\documentstyle[12pt]{article}
\begin{document}
\begin{titlepage}
\begin{centering}
{\large{\bf Classical moduli $O(\alpha')$ hair}}\\
\vspace{.7in}
{\bf P.Kanti and K.Tamvakis}\\
\vspace{.2in}%
Division of Theoretical Physics\\
Physics Department,University of Ioannina\\
Ioannina GR-451 10 ,GREECE\\
\vspace{.8in}
{\bf Abstract}\\
\vspace{.3in}
\end{centering}
{\noindent
We extend existing treatments of black hole solutions in String
Gravity to include moduli fields.We compute the external moduli
and dilaton hair, as well as of their associated axions,to
$O(\alpha')$ in the framework of the loop corrected superstring
effective action for a Kerr-Newman black hole background.}\\

\vspace{3.0in}
%
%\paragraph{}%

\par
\vspace{4mm}
\begin{flushleft}
IOA-317/95 \\
hep-th/9504031
\end{flushleft}
\end{titlepage}

%%%%%%%%%%%%%%%%
\par
Superstring $^{\cite{Green}}$ theory is our best existing candidate for a
consistent quantum theory of gravity which also has the prospect of
unification with all other interactions.Einstein's theory which has been
very successful as a classical theory of gravitation is incorporated in
this more general framework.However,Superstrings involve a
characteristic length of the order of the Planck scale and are expected
to lead to drastic modifications of the Einstein Action at short
distances.These modifications arise either due to the contribution of
the infinite tower of massive string modes,appearing as $\alpha'$
corrections,or due to quantum loop effects.An effective low energy
Lagrangian $^{\cite{Callan}}$ that incorporates the above, involving
only the massless string modes,can be derived from string theory using a
perturbative approach in both the string tension $\alpha'$ and the
string coupling.The relevant massless fields,apart from the graviton and
other gauge fields,are the dilaton,which plays the role of the
field-dependent string coupling that parametrizes the string loop
expansion, and the moduli fields that describe the size and the shape of
the internal compactification manifold.

In Einstein gravity,minimally coupled to other fields,the most general
black hole solution is described by the Kerr-Newman family of rotating
charged black hole solutions $^{\cite{Wald}}$.In agreement with the
``no-hair"theorem $^{\cite{Israel}}$ at the classical level the only
external fields present are those required by gauge invariance.A
qualitative new feature present in the superstring effective action is
the appearance of external field strength hair for the axion and dilaton
fields $^{\cite{Gibbons}}$ $^{-}$ $^{\cite{Olive}}$.The tree level
effective action has been calculated up to several orders in the
$\alpha'$-expansion.It turns out that there is no dependence on the
moduli fields at tree level.The one-loop corrections to gravitational
and gauge couplings have been calculated in the context of orbifold
compactifications of the heterotic superstring $^{\cite{Antoniadis}}$.It
has been shown that there are no moduli-dependent corrections to the
Einstein term while there are non trivial ${\cal {R}}^2$-contributions
appearing in a Gauss-Bonnet combination multiplied by a moduli-dependent
coefficient function.This term is subject to a non-renormalization
theorem which implies that all higher-loop moduli-dependent ${\cal
{R}}^2$-contributions vanish.It is interesting to note the existence of
singularity-free $^{\cite{Tamvakis}}$ solutions of the field equations
in a Friedmann-Robertson-Walker background depending crucially on the
presence of the Gauss-Bonnet term.

In the present short article we extend existing treatments
$^{\cite{Gibbons}}$ $^{-}$ $^{\cite{Olive}}$ of black hole solutions in
string gravity to iclude moduli fields. Our action is the low energy
effective action derived in the context of orbifold compactifications of
the heterotic superstring to one-loop and $\alpha'$-order.We compute to
$\alpha'$-order the moduli and dilaton-hair together with the
corresponding two axions hair.The result although expected from previous
existing investigations without the moduli fields serves to establish
even better the qualitatively  new features of string gravity in
contrast to Einstein gravity characterized by the ``no-hair" theorem.In
our action we have not introduced any potential for the above fields
although it is likely that in the full quantum string theory such a
potential and a (small) mass is generated through non-perturbative
effects.Nevertheless,if the black hole size,or the distance from the
black hole ,is small compared to their inverse mass the solutions found
are valid to a good approximation.

Let us consider the universal part of the effective action of any
four-dimensional heterotic superstring model  which describes the
dynamics of the graviton,gauge fields,the dilaton S and,for
simplicity,the common modulus field T.At the tree level and up to first
order in $\alpha'$ it takes the form
%%%%%%%%%%%%%%%
\begin{eqnarray}
S^{(o)} _{eff} & = & \int d^4 x \sqrt{-g} \left( {\frac {1} {2}} R
+ {\frac {|DS|^2} {(S+ \overline{S})^2}}
+ 3 {\frac {|DT|^2} {(T+\overline{T})^2}}
+{\frac {\alpha'} {8}} (ReS) ({\cal {R}}^2 _{GB}- F^{\mu
\nu}F_{\mu\nu}) \right.\nonumber\\
& + & \left.{\frac {\alpha'} {8}} (ImS) ({\cal {R}}
{\tilde{\cal{R}}}-F {\tilde{F}}) \right)
\end{eqnarray}
%%%%%%%%%%%%%%%
where
%%%%%%%%%%%%%%%
\begin{equation}
{\cal {R}}^2 _{GB} \equiv R_{\mu \nu \kappa \lambda} R^{\mu \nu \kappa
\lambda} - 4 R_{\mu \nu } R^{\mu \nu} + R^2
\end{equation}
%%%%%%%%%%%%%%%
and
%%%%%%%%%%%%%%%
\begin{equation}
{\cal{R}} {\tilde{\cal{R}}} \equiv {\cal{\eta}}^{\mu \nu \rho \sigma}
 R^{\kappa \lambda} _{\;\;\;\mu \nu} R_{\rho \sigma \kappa \lambda}
\end{equation}
%%%%%%%%%%%%%%%
\begin{equation}
F {\tilde{F}} \equiv {\cal{\eta}}^{\mu \nu \rho \sigma} F_{\mu \nu}
F_{\rho \sigma}
\end{equation}
%%%%%%%%%%%%%%%
In what follows we shall consider only the case of an Abelian gauge
field.Note that\footnote{$\epsilon^{oijk} = -\epsilon_{ijk}$}
${\cal{\eta}}^ {\mu \nu \rho \sigma} \equiv {\epsilon}^{\mu \nu \rho
\sigma} (-g)^{-1/2} $. We have chosen units such that $k \equiv \sqrt{8
\pi G_N } \equiv 1$.

The one-loop corrections give a modulus dependence to the quadratic
gravitational and gauge terms that are of the form
%%%%%%%%%%%%%%%
\begin{equation}
S^{(1)} _{eff} = \int
d^4x\sqrt{-g}\left(\alpha'\Delta(T,\overline{T}) {\cal{R}}^2 _{GB}
+\alpha' \Theta (T,\overline{T}) {\cal{R}} {\tilde{\cal{R}}}
+ \alpha' \hat{\Delta}(T,\overline{T}) F^{\mu \nu} F_{\mu
\nu}\right. \nonumber\\
+\left. \alpha' \hat{\Theta} (T,\overline{T}) F {\tilde{F}}
\right)
\end{equation}
%%%%%%%%%%%%%%%
The functions $\Delta(T,\overline{T})$,$\Theta (T,\overline{T})$,
$\hat{\Delta}(T,\overline{T})$ and $\hat{\Theta}
(T,\overline{T}) $ have been derived in ref.$[12]$ and depend
multiplicatively through a coefficient on the supermultiplet content of
the string model.Indroducing the notation
%%%%%%%%%%%%%%%
\begin{equation}
S \equiv ( e^{\phi} + i a)/{g^2} \quad ,\quad T \equiv e^{\sigma} + i b
\end{equation}
%%%%%%%%%%%%%%%
and reffering to $\phi$ as the dilaton, to $\sigma$ as the modulus,to
$a$ and b as the axions and to $g^2$ as the string coupling ,we can
write the effective one-loop, $O(\alpha')$ Langrangian as
%%%%%%%%%%%%%%%
\begin{eqnarray}
{\cal{L}}_{eff} & = & {\frac {1} {2}} R
+ {\frac {1} {4}} {({\partial} _{\mu} \phi)}^2
+ {\frac {1} {4}} e^{-2 \phi} {({\partial} _{\mu} a)}^2
+  {\frac {3} {4}} {({\partial} _{\mu}\sigma)}^2
+ {\frac {3} {4}} e^{-2\sigma} {({\partial} _{\mu} b)}^2 \nonumber\\
& + & \alpha' \left({\frac {e^{\phi}} {8 g^2}}
+ \Delta \right) {\cal{R}}^2 _{GB}
+  \alpha'\left({\frac {a} {8 g^2}}
+ \Theta \right){\cal{R}} {\tilde{\cal{R}}} \nonumber\\
& + & \alpha' \left(- {\frac {e^{\phi}} {8 g^2}}+
\hat{\Delta}\right) F^{\mu \nu} F_{\mu \nu} + \alpha' \left({-\frac
{a} {8 g^2}}  +\hat{\Theta} \right) F {\tilde{F}}
\end{eqnarray}
%%%%%%%%%%%%%%%
\newpage

The equations of motion resulting from $(6)$ are four equations for the
scalar and pseudoscalar fields
%%%%%%%%%%%%%%%
\begin{equation}
{\frac {1} {\sqrt{-g}}} \partial_{\mu}[\sqrt{-g} \partial^{\mu} \phi]
 = -e^{-2 \phi} (\partial_{\mu}a)^2  + {\frac {\alpha'} {4 g^2}}
e^{\phi}\left({\cal{R}}^2 _{GB}-F^{\mu \nu} F_{\mu \nu}\right)
\end{equation}
%%%%%%%%%%%%%%%
%
%%%%%%%%%%%%%%%
\begin{equation}
{\frac {1} {\sqrt{-g}}} \partial_{\mu}[\sqrt{-g} e^{-2 \phi}
\partial^{\mu} a]
= {\frac {\alpha'} {4g^2}}\left({\cal{R}} {\tilde{\cal{R}}}-
F {\tilde{F}}\right)
\end{equation}
%%%%%%%%%%%%%%%
%
%%%%%%%%%%%%%%%
\begin{eqnarray}
{\frac {1} {\sqrt{-g}}} \partial_{\mu}[\sqrt{-g} \partial^{\mu} \sigma]
&=& -e^{-2 \sigma}{({\partial} _{\mu} b)}^2 + {\frac{2 \alpha'}
{3}}\left( {\frac{\delta \Delta} {\delta \sigma}} \right) {\cal{R}}^2
_{GB} + {\frac {2 \alpha'}{3}} \left( {\frac {\delta \Theta}  {\delta
\sigma}}\right){\cal{R}}{\tilde{\cal{R}}} \nonumber \\
&+& {\frac{2 \alpha'}
{3}}\left( {\frac{\delta \hat{\Delta}} {\delta \sigma}} \right)F^{\mu
\nu}  F_{\mu \nu}  +
{\frac {2 \alpha'}{3}} \left( {\frac {\delta \hat{\Theta}}  {\delta
\sigma}}\right) F {\tilde{F}}
\end{eqnarray}
%%%%%%%%%%%%%%%
%
%%%%%%%%%%%%%%%
\begin{eqnarray}
{\frac {1} {\sqrt{-g}}} \partial_{\mu}[\sqrt{-g} e^{-2 \sigma}
\partial^{\mu}b] &=& {\frac{2 \alpha'} {3}}\left({\frac{\delta \Delta}
{\delta b}} \right) {\cal{R}}^2 _{GB} + {\frac {2 \alpha'}{3}} \left(
{\frac {\delta \Theta}  {\delta
b}}\right){\cal{R}}{\tilde{\cal{R}}} \nonumber \\
&+& {\frac{2 \alpha'}
{3}}\left( {\frac{\delta \hat{\Delta}} {\delta b}} \right)F^{\mu
\nu}  F_{\mu \nu}  +
{\frac {2 \alpha'}{3}} \left( {\frac {\delta \hat{\Theta}}  {\delta
b}}\right) F {\tilde{F}}
\end{eqnarray}
%%%%%%%%%%%%%%%
the gravitational equation\footnote{$\tilde{R}^{\mu \nu } _{\; \; \;
\kappa  \lambda}= {\cal{\eta}}^{\mu \nu \rho \sigma} R_{\rho \sigma
\kappa \lambda}$}
%%%%%%%%%%%%%%%
\begin{eqnarray}
& & R_{\mu \nu}-{\frac {1} {2}}g_{\mu \nu}R
+ \alpha'(g_{\mu \rho} g_{\nu\lambda}
+ g_{\mu \lambda} g_{\nu \rho}) {\cal{\eta}}^{\kappa \lambda \alpha
\beta}D_{\gamma}({{\tilde{R}}^{\rho \gamma}} _{\;\;\;\alpha \beta}
D_{\kappa} f_1) \nonumber \\
& &- 8 \alpha' D_{\rho}(\tilde{R}_{\mu \; \;
\nu} ^{\; \; \lambda \; \; \rho}  D_{\lambda} f_2) + 4 \alpha' f_3 (
F_{\mu} ^{\;\;\sigma} F_{\nu \sigma}-{\frac {1} {4}} g_{\mu \nu} F^{\rho
\sigma} F_{\rho \sigma}) =\nonumber \\& & -{\frac{1} {2}}(\partial_{\mu}
\phi) (\partial_{\nu} \phi) + {\frac{1} {4}} g_{\mu \nu}
(\partial_{\rho}\phi)^2 - {\frac{e^{-2 \phi}} {2}} (\partial_{\mu}a)
(\partial_{\nu}a) +{\frac {1} {4}}g_{\mu \nu} e^{-2 \phi}
(\partial_{\rho}a)^2\nonumber\\
& & -{\frac{3} {2}}
(\partial_{\mu}\sigma) (\partial_{\nu} \sigma) + {\frac {3} {4}} g_{\mu
\nu}(\partial_{\rho}\sigma)^2 -{\frac {3} {2}}e^{-2
\sigma}(\partial_{\mu}b) (\partial_{\nu}b) + {\frac {3} {4}} e^{-2
\sigma} g_{\mu \nu } (\partial_{\rho}b)^2
\end{eqnarray}
%%%%%%%%%%%%%%%
and the equation for the gauge field
%%%%%%%%%%%%%%%
\begin{equation}
{\frac {1} {\sqrt{-g}}} \partial_{\mu}[\sqrt{-g} f_3 F^{\mu \nu}] +
{\frac {1} {\sqrt{-g}}} \partial_{\mu}[\sqrt{-g} f_4 {\tilde{F}}^{\mu
\nu}]=0
\end{equation}
%%%%%%%%%%%%%%%
We  have introduced the functions
%%%%%%%%%%%%%%%
\begin{equation}
f_1 \equiv {\frac {e^{\phi}} {8 g^2}} + \Delta \quad , \quad
f_2 \equiv {\frac {a} {8 g^2}} + \Theta \quad , \quad
f_3 \equiv {-\frac {e^{\phi}} {8 g^2}} + \hat{\Delta} \quad , \quad
f_4 \equiv {-\frac {a} {8 g^2}} + \hat{\Theta}
\end{equation}
%%%%%%%%%%%%%%%

At this point we introduce the Kerr-Newman metric,the most general black
hole solution of the standard Einstein equation minimally coupled to an
Abelian gauge field.It is
\newpage
%%%%%%%%%%%%%%%
\begin{equation}
ds^2 = \left({\frac{{\rho}^2-2 M r+q^2} {{\rho}^2}}\right) dt^2 -
{\frac{{\rho}^2} {\Lambda}} dr^2 - {\rho}^2 d{\theta}^2 + {\frac {2
A{sin}^2\theta(2 M r-q^2)} {{\rho}^2}}dt d\varphi-{\frac{{sin}^2\theta}
{{\rho}^2}} {\Sigma}^2 d{\varphi}^2
\end{equation}
%%%%%%%%%%%%%%%
where  ${\rho}^2 \equiv r^2+A^2{cos}^2\theta$, $\Lambda \equiv r^2+A^2
-2Mr+q^2$ and ${\Sigma}^2 \equiv (r^2+A^2)^2-\Lambda A^2{sin}^2\theta$.
$A$ stands for the angular momentum per unit mass and $ q^2 = q_e^2 +
q_m^2$ for the total charge of the black hole.It will be shown very
shortly that the Kerr-Newman metric satisfies our gravitational
equation (12) to $O(\alpha')$.

Since,as we declared in the introduction,we plan to  determine solutions
to $O(\alpha')$ let us first obtain the zeroth order solutions for the
scalar and pseudoscalar fields.Introducing a rescaled axion field
$\partial_{\mu} \tilde{a} \equiv e^{-2 \phi} \partial_{\mu}a$ we can
write the dilatonic-axion equation of motion for a Kerr-Newman
background in the form
%%%%%%%%%%%%%%%
\begin{equation}
{\frac {\partial} {\partial r}}\left[(r^2-2Mr+A^2 +q^2){\frac{\partial
\tilde {a}} {\partial r}}\right] + {\frac {1}
{sin\theta}}{\frac{\partial} {\partial \theta}} \left[sin\theta
{\frac{\partial \tilde{a}} {\partial \theta}}\right] = 0
\end{equation}
%%%%%%%%%%%%%%%
It has a general solution of the form
%%%%%%%%%%%%%%%
\begin{equation}
\tilde{a} = \sum_{l=0}^{\infty} P_l (cos\theta)\left[A_l Q_l(z) + B_l
P_l(z)\right]
\end{equation}
%%%%%%%%%%%%%%%
where $z\equiv (r-M)/\sqrt{M^2-A^2-q^2}$. Imposing the black hole
boundary condition\footnote{The horizon of a Kerr-Newman black hole is
$r_H=M+\sqrt{M^2-A^2-q^2}$} $ r \rightarrow
r_H$  or  $ z\rightarrow 1$  forces us to require $ A_l =0 $, $\forall
l$.On the  other hand requiring finiteness at  $r\rightarrow \infty$
or  $z \rightarrow \infty$ forces us to set $ B_l=0$ , $\forall l\geq1$.
Thus, only the constant solution $ \tilde{a}= B_o$ is possible.Using
that,the dilaton equation reduces,to zeroth order,to the form
$D^2\phi=0$ which for the same reasons as in the case of the axion
$\tilde{a}$ leads us to the conclusion that to this order the dilaton is
a constant. Following the same procedure for the modulus and its
associated axion we also arrive at constant zeroth order values.

The gravitational equation to $O(\alpha')$ takes the form of the
minimal Einstein-Yang Mills equation
%%%%%%%%%%%%%%
\begin{equation}
R_{\mu \nu}-{\frac {1} {2}}g_{\mu \nu}R =
- 4 \alpha' (f_3)_{\phi_o,\sigma_o,b_o} ( F_{\mu} ^{\;\;\sigma} F_{\nu
\sigma}-{\frac {1} {4}} g_{\mu \nu} F^{\rho \sigma} F_{\rho \sigma})
\end{equation}
%%%%%%%%%%%%%%
At this point we can introduce for the gauge field the ansatz
%%%%%%%%%%%%%%%
\begin{equation}
{\cal{A}} = {\cal{A}}_{\mu} d x^{\mu} = {\frac {Q_e r} {{\rho}^2}} [dt-A
sin^2\theta d \varphi] + {\frac {Q_m cos\theta}  {{\rho}^2}} [A dt-(r^2 +
A^2 ) d \varphi]
\end{equation}
%%%%%%%%%%%%%%%
This ansatz describes a dyon since it possesses both an electric $Q_e$
and a magnetic $Q_m$ charge.Note that for vanishing $q^2$ the
gravitational equation can be satisfied with the standard Kerr
metric and there is no correction of $O(\alpha')$.In the opposite
case,there must be a $O(\alpha')$ correction from the gauge sector in
the metric.Thus,the charge $q^2$ should come out to be $O(\alpha')$. We
can easily derive the relation
%%%%%%%%%%%%%%%
\begin{equation}
q^2 = \alpha' Q^2 \left[ {\frac {e^{\phi_o}} {4 g^2}} -
2 \left(\hat{\Delta}(T,{\overline{T}}) \right)_{\sigma_o,b_o}\right]
\end{equation}
%%%%%%%%%%%%%%%
where $Q^2= Q_e^2+Q_m^2$.Due to the fact that the source terms are
already of $O(\alpha')$ any $O(\alpha')$ correction to the metric will
not affect the solution for the fields.Also in the limit $\alpha'
\rightarrow 0$,with $Q^2$ fixed,$q^2 \rightarrow 0$ and we can
use the Kerr metric (which follows from eq.(15) if we set
$q^2=0$) for our computations.For this metric we can calculate
%%%%%%%%%%%%%%%
\begin{equation}
{\cal{R}}{\tilde{\cal{R}}} =
{\frac {192M^2Arcos\theta(3r^2-A^2{cos}^2\theta)(r^2-3A^2{cos}^2\theta)}
{(r^2+A^2{cos}^2\theta)^6}}
\end{equation}
%%%%%%%%%%%%%%%
%
%%%%%%%%%%%%%%%
\begin{equation}
{\cal{R}}^2 _{GB} = {\frac
{48M^2(r^2-A^2{cos}^2\theta)[(r^2+A^2{cos}^2\theta)^2-16r^2A^2{cos}^2
\theta]} {(r^2+A^2{cos}^2\theta)^6}}
\end{equation}
%%%%%%%%%%%%%%%
Similarly,in terms of the ${\cal{A}}$ expression and the Kerr metric
we can calculate
%%%%%%%%%%%%%%%
\begin{eqnarray}
F^{\mu \nu} F_{\mu \nu}& = &-{\frac {2 (Q_e^2 - Q_m^2) [(r^2-A^2
cos^2\theta)^2-4 A^2 r^2 cos^2\theta]} {(r^2 + A^2
cos^2\theta)^4}}\nonumber \\
& -& {\frac {16 Q_e Q_m A rcos\theta (r^2-A^2 cos^2\theta) } {(r^2 + A^2
cos^2\theta)^4}}
\end{eqnarray}
%%%%%%%%%%%%%%%
%
%%%%%%%%%%%%%%%
\begin{eqnarray}
F {\tilde{F}}& =& - {\frac {16 (Q_e^2-Q_m^2) A r cos\theta (r^2-A^2
cos^2\theta)} {(r^2+A^2{cos}^2\theta)^4}}\nonumber \\
&+& {\frac {8 Q_e Q_m [(r^2-A^2 cos^2\theta)^2- 4 A^2 r^2 cos^2\theta]}
{(r^2+A^2{cos}^2\theta)^4}}
\end{eqnarray}
%%%%%%%%%%%%%%%

Following the same procedure as with the Kerr-Newman metric,we can see
that the zeroth-order solutions for the Kerr metric are still
constants.The solution for the dilaton,modulus and axion fields will be
derived from the $O(\alpha')$ equations $(8)-(11)$ setting the
zeroth-order solutions for the fields in the right-hand side.As a
result,the kinetic terms ${({\partial} _{\mu} a)}^2$ and
$(\partial_{\mu} b)^2$ vanish,the derivatives of
$\Delta$,$\Theta$,$\hat{\Delta}$ and $\hat{\Theta}$ are taken at the
point $(\sigma=\sigma_o$,$b=b_o)$ and the quadratic gravitational and
gauge terms are given from the expressions $(21)-(24)$.

In order to proceed and obtain the $O(\alpha')$ solutions we need the
static axisymmetric Green's function defined by the equation
%%%%%%%%%%%%%%%
\begin{equation}
{\frac {1} {\sqrt{-g}}} \partial_{\mu}\left[ \sqrt{-g} g^{\mu \nu}
\partial_{\nu}G(x-y)\right] = {\frac {\delta^{(3)}(x-y)} {\sqrt{-g}}}
\end{equation}
%%%%%%%%%%%%%%%
which for the Kerr metric becomes
%%%%%%%%%%%%%%%
\begin{equation}
{\frac {\partial} {\partial r}}\left[(r^2+A^2-2Mr){\frac {\partial G}
{\partial r}}\right]+ {\frac{1} {sin\theta}} {\frac {\partial}
{\partial \theta}}\left[sin\theta {\frac {\partial G} {\partial
\theta}}\right] = -\delta(r-r_o) \delta(cos\theta-cos{\theta}_o)
\delta(\varphi-{\varphi}_o)
\end{equation}
%%%%%%%%%%%%%%%
for a point source located at $r_o$ , $\theta_o$ , $ \varphi_o$.Demanding
finiteness at  $ r=r_H$ and at infinity we obtain
%%%%%%%%%%%%%%%
\begin{equation}
G(r,\theta,\varphi; r_o,\theta_o,\varphi_o) = \sum_{l=0}^{\infty} R_l
(r,r_o)  P_l (cos\gamma)
\end{equation}
%%%%%%%%%%%%%%%
with
%%%%%%%%%%%%%%%
\begin{equation}
cos\gamma \equiv cos\theta cos\theta_o + sin\theta sin\theta_o
cos(\varphi-\varphi_o)
\end{equation}
%%%%%%%%%%%%%%%
and
%%%%%%%%%%%%%%%
\begin{eqnarray}
R_l  (r,r_o)& = & - {\frac {(2l+1)} { 4\pi \sqrt{M^2-A^2}}}  \left[ P_l
\left({\frac {(r_o-M)} {\sqrt{M^2-A^2}}} \right)  Q_l \left(
{\frac {(r-M)} {\sqrt{M^2-A^2}}} \right)  \theta(r-r_o) \right. \nonumber
\\ & + & \left.  P_l\left( {\frac {(r-M)} {\sqrt{M^2-A^2}}} \right) Q_l
\left({\frac{(r_o-M)} {\sqrt{M^2-A^2}}}\right)  \theta( r_o-r) \right]
\end{eqnarray}
%%%%%%%%%%%%%%%

Using the Green's function we can write  the external dilaton solution as
%%%%%%%%%%%%%%%
\begin{equation}
\phi(r,\theta,\varphi) = \int_{r_H}^{\infty} dr_o \int_{-1}^1 dcos{
\theta}_o \int_{0}^{2\pi}d{\varphi}_o (r_o^2 + A^2 {cos}^2\theta_o)
G(r,\theta,\varphi;r_o,\theta_o,\varphi_o) {\cal{J}}(r_o,\theta_o,
\varphi_o)
\end{equation}
%%%%%%%%%%%%%%%
where the source ${\cal{J}}$ is the right hand side of equation $(8)$.
Similar expressions hold for the rest of the scalar and pseudoscalar
fields\footnote{$\partial_{\mu} \tilde{b}=e^{-2 \sigma} \partial_{\mu}
b$}  $\sigma$,$\tilde{a}$,$\tilde{b}$ in terms of the corresponding
source terms.It is straightforward but tedious to obtain the $O(\alpha')$
expressions for these fields.The $O(\alpha')$ fields are
%%%%%%%%%%%%%%
\begin{eqnarray}
\phi(r,\theta)& =& \phi_o - {\frac { \alpha' e^{\phi_o} } {g^2}}  \left[
  {\frac {1} { A^2}} ln\left({\frac {r-M+\sqrt{M^2-A^2}}
{\sqrt{A^2+r^2}}}\right) + {\frac {2 A^2-M^2} {2 A^3 M}}\left(
{\frac{\pi} {2}} - Arctan{\frac {r} {A}}\right) \right. \nonumber\\&+&
\left.  {\frac {M r} {(A^2+r^2)^2}} + {\frac {2
A^2+M r}  {2 A^2 (A^2+r^2)}} + {\frac
{Q_e^2-Q_m^2} {4 A M}}\left({\frac{\pi}  {2}} - Arctan{\frac {r}
{A}})\right. \right]P_o(cos\theta)\nonumber \\& + & {\frac {3 \alpha'
 Q_e Q_m e^{\phi_o}} {4 A M g^2}} \left({\frac {r} {A}} Arctan{\frac {A}
{r}} -1 \right) P_1(cos\theta)+...
\end{eqnarray}
%%%%%%%%%%%%%%
%
%%%%%%%%%%%%%%
\begin{eqnarray}
\tilde{a}(r,\theta) & = & \tilde{a}_o
+ {\frac {\alpha' Q_e Q_m} {A M g^2}} \left( {\frac{\pi} {2}}
- Arctan{\frac {r} {A}}\right) P_o(cos\theta)
+ \left \{ {\frac { 3 \alpha' (Q_e^2-Q_m^2)} {2 A M g^2}} \left(
{\frac {r} {A}} Arctan{\frac {A} {r}} -1 \right) \right. \nonumber \\
& - & \left. {\frac {6 \alpha' A} {g^2 (M^2-A^2) }} \left[ (r-M) \left(
{\frac {A^2- M^2} {A^4}} ln \left( {\frac {r-M+\sqrt{M^2-A^2}}
{\sqrt{A^2+r^2}}} \right) \right. \right. \right. \nonumber \\
& +& \left. \left. \left.{\frac {2 A^2+Mr-M^2} {2 A^2  (A^2+r^2)}}
+{\frac {A^2-M^2 }  {A^3 M}} \left( {\frac {\pi} {2}}
- Arctan{\frac {r} {A}}\right) \right)
+ {\frac {A^2 \sqrt{M^2-A^2}} { M r_H ^3}} \right. \right. \nonumber \\
& +& \left.  \left. {\frac {3 M-4 r_H} {2 r_H ^2}}
+{\frac {M (M^2+r^2)} { (A^2+r^2)^2}} \right] \right\}P_1(cos\theta)
+...
\end{eqnarray}
%%%%%%%%%%%%%%
%
%%%%%%%%%%%%%%
\begin{eqnarray}
\sigma(r,\theta) & = & \sigma_o
+ \left \{ - {\frac {8\alpha'} {3}} \left( {\frac
{\partial \Delta} {\partial \sigma}} \right)_{\sigma_o,b_o} \left[
{\frac {1} { A^2}} ln \left( {\frac {r-M+\sqrt{M^2-A^2}}
{\sqrt{A^2+r^2}}} \right)
 \right. \right. \nonumber\\
&+& \left. \left.{\frac {2 A^2-M^2} {2 A^3 M}} \left(
{\frac{\pi} {2}} - Arctan{\frac {r} {A}} \right) +{\frac {2 A^2+M r}
{2 A^2 (A^2+r^2)}}+{\frac {M r} {(A^2+r^2)^2}} \right] \right.
\nonumber \\
&+& \left. \left( {\frac {\partial \hat{\Delta}} {\partial
\sigma}} \right)_{\sigma_o,b_o}  {\frac {2 \alpha'(Q_e^2-Q_m^2)}
{3 A M}} \left({\frac{\pi} {2}} -Arctan{\frac {r} {A}}\right) \right.
\nonumber \\
&-&\left. \left( {\frac {\partial \hat{\Theta}}  {\partial
\sigma}} \right)_{\sigma_o,b_o} {\frac  {8 \alpha' Q_e Q_m} {3 A M }}
\left( {\frac{\pi} {2}} - Arctan{\frac {r} {A}}\right) \right\} P_o
(cos\theta) \nonumber \\
& - & \left\{ \left( {\frac{\partial \Theta} {\partial \sigma}}
\right)_{\sigma_o,b_o} {\frac {48 \alpha' A} { 3 (M^2-A^2)}} \left[
(r-M) \left( {\frac {A^2- M^2} {A^4}} ln \left( {\frac
{r-M+\sqrt{M^2-A^2}} {\sqrt{A^2+r^2}}} \right)\right. \right. \right.
\nonumber \\
& +&\left. \left. \left.{\frac {2 A^2+Mr-M^2} {2 A^2 (A^2+r^2)}}
+{\frac {A^2-M^2} {A^3 M}} \left( {\frac {\pi} {2}} -
Arctan{\frac {r} {A}}\right) \right)
+ {\frac {A^2 \sqrt{M^2-A^2}} { M r_H ^3}}\right.\right. \nonumber \\
& +&  \left. \left. {\frac {3 M-4 r_H} {2 r_H ^2}}
+{\frac {M (M^2+r^2)} { (A^2+r^2)^2}} \right]
+ \left( {\frac {\partial \hat{\Delta}} {\partial \sigma}}
\right)_{\sigma_o,b_o} {\frac {2 \alpha' Q_e Q_m} { A M }} \left({\frac
{r} {A}} Arctan{\frac {A} {r}} -1 \right) \right.  \nonumber \\
& + & \left. \left( {\frac {\partial \hat{\Theta}}  {\partial \sigma}}
\right)_{\sigma_o,b_o} {\frac {4 \alpha'(Q_e^2-Q_m^2)} {3 A M }}
\left({\frac {r} {A}}  Arctan{\frac {A} {r}} -1 \right)
\right \}P_1(cos\theta)  +...
\end{eqnarray}
%%%%%%%%%%%%%%
%
%%%%%%%%%%%%%%
\begin{eqnarray}
\tilde{b}(r,\theta) & = &\tilde{b}_o + \left \{ - {\frac {8\alpha'} {3}}
\left( {\frac {\partial \Delta} {\partial b}} \right)_{\sigma_o,b_o}
\left[ {\frac {1} { A^2}} ln \left( {\frac {r-M+\sqrt{M^2-A^2}}
{\sqrt{A^2+r^2}}} \right)
 \right. \right. \nonumber\\
&+& \left. \left.{\frac {2 A^2-M^2} {2 A^3 M}} \left(
{\frac{\pi} {2}} - Arctan{\frac {r} {A}} \right) +{\frac {2 A^2+M r}
{2 A^2 (A^2+r^2)}}+{\frac {M r} {(A^2+r^2)^2}} \right] \right.
\nonumber \\
&+& \left. \left( {\frac {\partial \hat{\Delta}} {\partial
b}} \right)_{\sigma_o,b_o}  {\frac {2 \alpha'(Q_e^2-Q_m^2)}
{3 A M}} \left({\frac{\pi} {2}} -Arctan{\frac {r} {A}}\right) \right.
\nonumber \\
&-&\left. \left( {\frac {\partial \hat{\Theta}}  {\partial
b}} \right)_{\sigma_o,b_o} {\frac  {8 \alpha' Q_e Q_m} {3 A M }}
\left( {\frac{\pi} {2}} - Arctan{\frac {r} {A}}\right) \right\} P_o
(cos\theta) \nonumber \\
& - & \left\{ \left( {\frac{\partial \Theta} {\partial b}}
\right)_{\sigma_o,b_o} {\frac {48 \alpha' A} { 3 (M^2-A^2)}} \left[
(r-M) \left( {\frac {A^2- M^2} {A^4}} ln \left( {\frac
{r-M+\sqrt{M^2-A^2}} {\sqrt{A^2+r^2}}} \right)\right. \right. \right.
\nonumber \\
& +&\left. \left. \left.{\frac {2 A^2+Mr-M^2} {2 A^2 (A^2+r^2)}}
+{\frac {A^2-M^2} {A^3 M}} \left( {\frac {\pi} {2}} -
Arctan{\frac {r} {A}}\right) \right)
+ {\frac {A^2 \sqrt{M^2-A^2}} { M r_H ^3}}\right.\right. \nonumber \\
& +&  \left. \left. {\frac {3 M-4 r_H} {2 r_H ^2}}
+{\frac {M (M^2+r^2)} { (A^2+r^2)^2}} \right]
+ \left( {\frac {\partial \hat{\Delta}} {\partial b}}
\right)_{\sigma_o,b_o} {\frac {2 \alpha' Q_e Q_m} { A M }} \left({\frac
{r} {A}} Arctan{\frac {A} {r}} -1 \right) \right.  \nonumber \\
& + & \left. \left( {\frac {\partial \hat{\Theta}}  {\partial b}}
\right)_{\sigma_o,b_o} {\frac {4 \alpha'(Q_e^2-Q_m^2)} {3 A M }}
\left({\frac {r} {A}}  Arctan{\frac {A} {r}} -1 \right)
\right \}P_1(cos\theta)  +...
\end{eqnarray}
%%%%%%%%%%%%%%

  The leading modulus and $\tilde{b}$-hair behaviour is that of a
monopole term analogous to the dilaton and $\tilde{a}$-hair.This is
evident from the slow rotation limit of the  dilaton solution
%%%%%%%%%%%%%%
\begin{eqnarray}
\phi(r,\theta)& = &\phi_o + {\frac {\alpha' e^{\phi_o}} {4 g^2}} \left\{
\left[-{\frac {2} {M r}} (1+{\frac {M} {r}} + {\frac {4 M^2} {3 r^2}}) +
{\frac {A^2} {2 M^3 r}} ( {\frac {1} {2}}+ {\frac {M} {r}} +
\right.\right. \nonumber\\ & + & \left.\left.{\frac {12 M^2} {3
r^2}} + {\frac {6 M^3} {r^3}} + {\frac {64 M^4} {5 r^4}} ) - {\frac
{(Q_e^2-Q_m^2)} {M r}} +...\right] P_o (cos\theta)\right.\nonumber \\
&-&\left.{\frac {A Q_e Q_m} { M r^2 }}
P_1(cos\theta)+...\right\}
\end{eqnarray}
%%%%%%%%%%%%%%
and the $\tilde{a}$-hair solution
%%%%%%%%%%%%%%
\begin{eqnarray}
\tilde{a}(r,\theta)& = &\tilde{a}_o +{\frac {\alpha'} {g^2}}\left\{
\left[{\frac  { Q_e Q_m} {M r}} + ...\right] P_o (cos\theta)
+\left[- {\frac { (Q_e^2-Q_m^2) A^2} {2 M r^2 }}\right.\right.
\nonumber \\ & - & \left.\left. {\frac {5 A} {4 M r^2 }} \left(
1+{\frac {2 M} {r}} + {\frac {18 M^2} {5 r^2}}\right)+... \right]
P_1(cos\theta)+...\right\}
\end{eqnarray}
%%%%%%%%%%%%%%
Note that the coefficient
functions$^{\cite{Antoniadis}}$ $\Delta$,$\Theta$,$\hat{\Delta}$ and
$\hat{\Theta}$ could be such that they have an extremum at the self-dual
point $\sigma_o=b_o=0$.Perturbing around the self-dual solution
leads  to vanishing modulus hair to $O(\alpha')$.The infinite continuum
of non-zero $\sigma_o$,$b_o$ values allows for the non-vanishing
modulus and $\tilde{b}$-axion hair given by $(33)$ and $(34)$.

Although the existence of non-trivial dilaton,moduli and axion fields
outside a Kerr-Newman black hole seems to violate the letter of the
``no-hair theorem" it does not violate the spirit since the solution is
uniquelly characterized by mass,charge and angular momentum.In the
terminology introduced by S.Coleman,J.Preskill and F.Wilczek
$^{\cite{Coleman}}$ the external moduli and dilaton hair are examples of
``secondary " hair. \newpage

\noindent
{\bf Acknowledgments}\\ \\
We thank I.Antoniadis,N.Mavromatos and J.Rizos for illuminating
discussions.

\topmargin=.2cm
%%%%%%%%%%%%%%%%%%%%%%%%%%%%%

\newpage


\begin {thebibliography}{99}
\bibitem{Green}For a review see : M.Green , J.Schwarz and  E.Witten,
``Superstring Theory "\\ (Cambridge U.P., Cambridge,1987).
%%
\bibitem{Callan} C.G.Callan, D.Friedan, E.J.Martinec and M.J.Perry,
Nucl.Phys.B262 (1985)543 ; B278 (1986)78.\\
E.S.Fradkin and A.A.Tseytlin, Phys.Lett.B158(1985)316; Nucl.Phys.B262
(1985)1.\\
A.Sen,Phys.Rev. D32 (1985)2142; Phys.Rev.Lett.55(1985)1846.
%%
\bibitem{Wald} R.M.Wald ,``General Relativity" (University of Chicago
Press, Chicago,1984) and references therein.
%%
\bibitem{Israel}For a review see:``General Relativity, an Einstein
Centennary Survey"\\ edit.S.W.Hawking and W.Israel(Cambridge
U.P.,Cambridge,1979).\\
W.Israel,Phys.Rev.164(1967)331.\\
B.Carter,Phys.Rev.Lett.26(1971)331.\\
R.Price,Phys.Rev.D5(1972)2419,2439.\\
J.D.Bekenstein,Phys.Rev.D5(1972)1239,2403.\\
C.Teitelboim,Phys.Rev.D5(1972)2941.\\
D.C.Robinson,Phys.Rev.Lett.34(1975)405.\\
R.M.Wald,Phys.Rev.Lett.26(1971)1653.\\
E.D.Fackerell and J.R.Ipser,Phys.Rev.D5(1972)2455.\\
S.L.Adler and R.B.Pearson,Phys.Rev.D18(1978)2798.
%%
\bibitem{Gibbons} G.W.Gibbons,Nucl.Phys.207(1982)337.\\
G.W.Gibbons and K.Maeda,Nucl.Phys.B298(1988)741.
%%
\bibitem{Perry} C.G.Callan,R.C.Myers and M.J.Perry,Nucl.Phys.B311(1988/
89)673.
%%
\bibitem{Campbell} B.Campbell,M.Duncan,N.Kaloper and
K.A.Olive,Phys.Lett.B251(1990)34.
%%
\bibitem{Garfinkle} D.Garfinkle,G.T.Horowitz and
A.Strominger,Phys.Rev.D43(1991)3140.
%%
\bibitem{Kaloper} B.Campbell,N.Kaloper and K.A.Olive,Phys.Lett.B263
(1991)364.
%%
\bibitem{Shapere} A.Shapere,S.Trivedi and F.Wilczek,Mod.Phys.Lett.A6
(1991)2677.
%%
\bibitem{Olive} B.Campbell,N.Kaloper and K.A.Olive,Phys.Lett.B285
(1992)199.
%%
\bibitem{Antoniadis} I.Antoniadis,E.Gava and
K.S.Narain,Phys.Lett.B283(1992)209;\\ Nucl.Phys.B383(1992)93.
%%
\bibitem{Tamvakis} I.Antoniadis,J.Rizos and K.Tamvakis,Nucl.Phys.B415
(1994)497.
%%
\bibitem{Coleman} S.Coleman,J.Preskill and F.Wilczek,Nucl.Phys.B378
(1992)175.
%%
\end{thebibliography}
\end{document}